\documentclass[orivec]{llncs}

\usepackage{latexsym,amssymb,amsmath}
\usepackage{url}
\usepackage[defblank]{paralist}
\usepackage{tikz,subfigure}
\usetikzlibrary{arrows,shapes,snakes,automata,backgrounds,petri,positioning,decorations.pathmorphing,fit,positioning,calc,backgrounds}
\usepackage{todonotes}
\usepackage{xspace}

\usepackage[thmmarks, amsmath]{ntheorem}
\theoremstyle{plain}
\theorembodyfont{\normalfont}
\theoremseparator{.}
\theoremsymbol{\ensuremath{\square}}
\newtheorem{notion}{Definition}

\theoremsymbol{\ensuremath{\qedboxfull}}
\theorembodyfont{\normalfont\small}
\newtheorem{example}{Example}

\newcommand{\dbnet}{db-net\xspace}
\newcommand{\dbnets}{db-nets\xspace}

\newcommand{\dbn}{\mathcal{B}}

\newcommand{\set}[1]{\{#1\}}
\newcommand{\tup}[1]{\langle#1\rangle}
\newcommand{\mult}[1]{#1^\oplus}

\newcommand{\true}{\mathsf{true}}
\newcommand{\false}{\mathsf{false}}

\newcommand{\qedboxfull}{\vrule height 5pt width 5pt depth 0pt}

\newcommand{\funsym}[1]{\mathtt{#1}}
\newcommand{\setsym}[1]{\mathit{#1}}

\newcommand{\R}{\mathcal{R}}
\newcommand{\E}{\mathcal{E}}

\newcommand{\I}{\mathcal{I}}

\newcommand{\schema}{\R}

\newcommand{\adom}[2][]{\setsym{Adom}_{#1}(#2)}

\newcommand{\free}[1]{\setsym{Free}(#1)}

\newcommand{\subst}{\theta}

\newcommand{\relname}[1]{\ensuremath{\mathit{#1}}\xspace} 
\newcommand{\cname}[1]{\ensuremath{\mathtt{#1}}\xspace} 

\newcommand{\pname}[1]{\ensuremath{\mathit{#1}}\xspace} 

\newcommand{\action}{\alpha}
\newcommand{\doact}[2]{\funsym{apply}(#1,#2)}

\newcommand{\typename}[1]{\mathbf{#1}}
\newcommand{\types}{\mathfrak{D}}
\newcommand{\type}{\mathcal{D}}

\newcommand{\dom}{\Delta}
\newcommand{\sigp}{\Gamma}
\newcommand{\reals}{\mathbb{R}}

\newcommand{\ints}{\mathbb{Z}}
\newcommand{\strings}{\mathbb{S}}
\newcommand{\vartype}{\funsym{type}}

\newcommand{\ans}{\ensuremath{\mathit{ans}}\xspace}

\newcommand{\typed}[2]{\ensuremath{#1\! : \! #2}}

\newcommand{\ptype}{\funsym{type}}
\newcommand{\fod}[1][\types]{\textnormal{\texttt{FO}(\ensuremath{#1})}\xspace}
\newcommand{\qent}[3]{#1,#2\models #3}
\newcommand{\qnent}[3]{#1,#2\not\models #3}
\newcommand{\cset}{\mathcal{\E}}	

\newcommand{\pers}{\mathcal{P}}

\newcommand{\datq}[3]{#1(#2)\textsc{:-\,}#3}
\newcommand{\actions}{\mathcal{A}}
\newcommand{\queries}{\mathcal{Q}}
\newcommand{\dl}{\mathcal{L}}

\newcommand{\aop}[1]{\mathtt{#1}}
\newcommand{\aname}[1]{{#1}{\cdot}\aop{name}}
\newcommand{\apar}[1]{{#1}{\cdot}\aop{params}}
\newcommand{\aadd}[1]{{#1}{\cdot}\aop{add}}
\newcommand{\adel}[1]{{#1}{\cdot}\aop{del}}
\newcommand{\actname}[1]{\textsc{#1}}

\newcommand{\varsin}[1]{\setsym{Vars}(#1)}
\newcommand{\inflow}{F_{in}}
\newcommand{\outflow}{F_{out}}
\newcommand{\rbflow}{F_{rb}}

\newcommand{\colors}{\Sigma}

\newcommand{\ibind}[2]{{#1}^\oplus(#2)}

\newcommand{\places}{P}
\newcommand{\cplaces}{P_c}
\newcommand{\vplaces}{P_v}
\newcommand{\coloring}{\funsym{color}}
\newcommand{\quass}{\funsym{query}}
\newcommand{\guass}{\funsym{guard}}
\newcommand{\nuvarset}{\Upsilon_{\types}}
\newcommand{\vars}{\mathcal{X}_{\types}}

\newcommand{\transitions}{T}

\newcommand{\guards}[1]{\mathbb{F}_{#1}}

\newcommand{\indact}[2]{\funsym{act}_{#2}(#1)}

\newcommand{\tuples}[1]{\Omega_{#1}}

\newcommand{\invars}[1]{\setsym{InVars}(#1)}
\newcommand{\outvars}[1]{\setsym{OutVars}(#1)}
\newcommand{\freshvars}[1]{\setsym{FreshVars}(#1)}

\newcommand{\aass}{\funsym{act}}

\newcommand{\cl}{\mathcal{N}}

\newcommand{\mode}{\sigma}

\newcommand{\varset}{\mathcal{V}_{\types}}

\newcommand{\enabled}[3]{#1 [#2,#3\rangle}
\newcommand{\fire}[4]{\enabled{#1}{#2}{#3}#4}

\newcommand{\marking}{\mathit{m}}

\newcommand{\dbstate}[2]{\tup{#1,#2}}
\newcommand{\tsys}[1]{\Gamma_{#1}}
\newcommand{\states}{S}
\newcommand{\istate}{s_0}
\newcommand{\state}{s}

\newcommand{\trans}{\rightarrow}

\tikzstyle{place}=[circle,thick,draw=black,fill=white,minimum size=7mm,font=\fontsize{9}{144}\selectfont]
\tikzstyle{transition}=[rectangle,thick,draw=black,fill=gray!20,minimum size=7mm]
\tikzstyle{enabledtransition}=[rectangle,very thick,draw=green!75,fill=green!20,minimum size=7mm]
 \tikzstyle{container}=[rectangle,rounded corners,very thick,draw=black!75,fill=black!20,minimum height=7mm,minimum width=14mm]
\tikzstyle{rplace}=[circle,ultra thick,draw=violet!75,fill=violet!20,minimum size=7mm]

\tikzset{state/.style={rectangle,rounded corners,draw=black,  thick, inner sep=2pt, text centered},}
\tikzset{statet/.style={rectangle,rounded corners,draw=black,  thick, inner sep=2pt, text centered,minimum size=1.2cm},}

\def\dbicon#1#2#3{
    \node at #1 [cylinder, shape border rotate=90, draw=white,fill=black,minimum height=#2,minimum width=#3,yshift=-.1cm] {};
}

\def\viewplace#1#2#3{
\begin{scope}[shift={#2}]
    \node (#1) at (0,0) [place,draw,label={#3}] {};
    \node at (0,0) [cylinder, shape border rotate=90, draw=white,fill=black,minimum height=.3cm,minimum width=.4cm,yshift=-.1cm] {};
\end{scope}
}

\makeatletter
\pgfarrowsdeclare{X}{X}
{
  \pgfutil@tempdima=0.3pt%
  \advance\pgfutil@tempdima by.25\pgflinewidth%
  \pgfutil@tempdimb=5.5\pgfutil@tempdima\advance\pgfutil@tempdimb by.5\pgflinewidth%
  \pgfarrowsleftextend{+-\pgfutil@tempdimb}
  \pgfarrowsrightextend{0pt}
}
{
  \pgfutil@tempdima=0.3pt%
  \advance\pgfutil@tempdima by.25\pgflinewidth%
  \pgfsetdash{}{+0pt}
  \pgfsetroundcap
  \pgfsetmiterjoin
  \pgfpathmoveto{\pgfqpoint{-5.5\pgfutil@tempdima}{-6\pgfutil@tempdima}}
  \pgfpathlineto{\pgfqpoint{5.5\pgfutil@tempdima}{6\pgfutil@tempdima}}
  \pgfpathmoveto{\pgfqpoint{-5.5\pgfutil@tempdima}{6\pgfutil@tempdima}}
  \pgfpathlineto{\pgfqpoint{5.5\pgfutil@tempdima}{-6\pgfutil@tempdima}}
  \pgfusepathqstroke 
}
\makeatother

\makeatletter
\pgfarrowsdeclare{x}{x}
{
  \pgfutil@tempdima=0.3pt%
  \advance\pgfutil@tempdima by.25\pgflinewidth%
  \pgfutil@tempdimb=5.5\pgfutil@tempdima\advance\pgfutil@tempdimb by.5\pgflinewidth%
  \pgfarrowsleftextend{+-\pgfutil@tempdimb}
  \pgfarrowsrightextend{0pt}
}
{
  \pgfutil@tempdima=0.3pt%
  \advance\pgfutil@tempdima by.25\pgflinewidth%
  \pgfsetdash{}{+0pt}
  \pgfsetroundcap
  \pgfsetmiterjoin
  \pgfpathmoveto{\pgfqpoint{-3.5\pgfutil@tempdima}{-4\pgfutil@tempdima}}
  \pgfpathlineto{\pgfqpoint{3.5\pgfutil@tempdima}{4\pgfutil@tempdima}}
  \pgfpathmoveto{\pgfqpoint{-3.5\pgfutil@tempdima}{4\pgfutil@tempdima}}
  \pgfpathlineto{\pgfqpoint{3.5\pgfutil@tempdima}{-4\pgfutil@tempdima}}
  \pgfusepathqstroke 
}
\makeatother

\title{DB-Nets: on The Marriage of\\ Colored Petri Nets and Relational Databases}
\titlerunning{DB-nets}
\author{Marco Montali \and Andrey Rivkin}
\authorrunning{Montali, Rivkin}
\institute{
 Free University of Bozen-Bolzano,
 Piazza Domenicani 3, 39100 Bolzano, Italy\\
 \email{montali,rivkin@inf.unibz.it}
}

\begin{document}

\maketitle

\begin{abstract}
The integrated management of business processes and master data is being increasingly considered as a fundamental problem, by both the academia and the industry. In this position paper, we focus on the foundations of the problem, arguing that contemporary approaches struggle to find a suitable equilibrium between data- and process-related aspects. We then propose \dbnets, a new formal model that balances such two pillars through the marriage of colored Petri nets and relational databases. We invite the research community to build on this model, discussing  its potential in modeling, formal verification, and simulation.
\end{abstract}

\section{Introduction}
In contemporary organizations, the integrated management of business processes (BPs) and master data is being increasingly considered as a fundamental problem, both by academia and the industry.
From the practical point of view, it has been widely recognized that the traditional isolation between process and data management induces a  fragmentation and redundancies in the organizational structure and its underlying IT solutions, with experts and tools solely centered around data, and others only focusing on process management \cite{Rich10,Dum11,Reic12}. This isolation falls short, especially when it comes to knowledge-intensive and human-empowered processes \cite{Hull08,AaWG05,KuWR11}. 
Foundational research has witnessed a similar trend, with completely separate fields of research either focused on the foundations of data management, or the principles underlying dynamic concurrent systems, database theory and Petri net theory being the two most prominent representatives within such fields. 

To answer the increasing demand of integrated models holistically tackling the dynamics of a complex domain and the manipulation of data, a plethora of approaches has emerged in the last decade. Such approaches can be classified in two main groups, again implicitly reflecting the process-data dichotomy. 
 A first series of approaches comes from Petri nets, the reference formalism to represent the control-flow of BPs, with the purpose of increasing their awareness of data. All such models are more or less directly inspired by a class of high-level Petri nets called Colored Petri nets (CPNs) \cite{JeK09,AaS11}, where colors abstractly account for data types, and where control threads (i.e., tokens) progressing through the net carry data conforming to colors. In their original formulation, though, CPNs must be severely restricted when it comes to their formal analysis, in particular requiring color domains to be finite, and in principle allowing one to get rid of the data via propositionalization. Several data-aware fragments of high-level Petri nets that are amenable to formal analysis even in the case of infinite color domains have consequently been studied, ranging from nets where tokens carry single data values (as in data- and $\nu$-nets \cite{Las16,RVFE11}), to nets where tokens are associated to more complex data structures such as nested relations \cite{HKSTVdB08}, nested terms \cite{TriS16}, or XML documents \cite{BHM15}.
The common, main issue of al such approaches is that data are still subsidiary to the control-flow dimension: data elements ``locally" attached to tokens, without being connected to each other by a global, end-to-end data model. In this light, CPN-based approaches      
%
%
naturally support the key notion of process instance (or case) through that of token, together with the key notion of case attributes \cite{RHE05}. However, they do not lend themselves to modeling, querying, updating, and ultimately reasoning on persistent, relational data, like those typically maintained inside an enterprise information system. For this reason, they are unable to suitably formalize concrete BP execution engines, which all provide support for explicitly interconnecting a BP with an underlying relational persistent storage \cite{DDG16}.

The second group of foundational approaches to data-aware processes has emerged at the intersection of database theory, formal methods and conceptual modeling, and specularly mirrors the advantages and lacks of CPN-based solutions. Such proposals go under the umbrella term of data-centric approaches \cite{CaDM13}, and gained momentum during the last decade, in particular due to the development of the business artifact paradigm \cite{CohnH09}, which also lead to concrete languages and implementations 	\cite{DaHV13,KuWR11}. 
The common denominator of all such approaches is that processes are centered around an explicit, persistent data component maintaining information about the domain of interest, and possibly capturing also its semantics in terms of classes, relations, and constraints. Atomic tasks induce CRUD (create-ready-update-delete) operations over the data component, in turn supporting the evolution of the master data maintained therein.
Proposals then differ in terms of   
the adopted data model  (e.g., relational, tree-shaped, graph-structured), and on the nature of information (e.g., whether it is complete or not). For example, \cite{BCDD13,DHPV09} focus on relational databases, while \cite{AbSV09} refers to XML.
The main downside of data-centric process models is that they disregard an explicit representation of how atomic tasks have to be sequenced over time, only implicitly representing the control flow via (event-)condition-action rules \cite{DHPV09,DaHV13,BCDD13}. 

In this position paper, we argue that the lack of equilibrium between data- and process-related aspects in the aforementioned proposals is a major obstacle towards modeling, verifying, monitoring, mining, and ultimately \emph{understanding} data-aware business processes. We believe that this can only be achieved by better balancing such two pillars. This, in turn, calls for the development of further modeling abstractions tailored to 
	establishing more intimate, synergic connections between CPNs and data-centric approaches. To the best of our knowledge, the only existing proposal that makes an effort in this direction is \cite{DDG16}. However, it employs workflow nets \cite{Aal97} for capturing the process control flow, without leveraging the advanced capabilities of CPNs. 
Taking inspiration from \cite{DDG16}, we then propose \emph{\dbnets}, a new, balanced formal model for data-aware processes, rooted in CPNs and relational databases. We rigorously describe the abstractions offered by the model, and formalize its execution semantics. 
We finally invite the research community to build on this new model, discussing its potential along three subjects: modeling, verification, and simulation.

\section{The DB-Net Model}

\definecolor{dbcolor}{HTML}{F39C12}
\definecolor{datacolor}{HTML}{FAD7A0}
\definecolor{activedatacolor}{HTML}{85C1E9}
\definecolor{viewcolor}{HTML}{5499C7}
\definecolor{actioncolor}{HTML}{2471A3}
\definecolor{netcolor}{HTML}{D576AE}

\begin{figure}[t]
\centering
\resizebox{.75\hsize}{!}{
\begin{tikzpicture}[x=1cm,y=1.3cm,>=triangle 60,thick]
\tikzstyle{elem} = [rectangle,rounded corners=5pt,minimum height=.7cm,minimum width=1.5cm,draw]
\tikzstyle{back} = [rectangle,rounded corners=10pt,minimum height=1cm,minimum width=8cm,anchor=west]
\node (db) at (0,-.25) [cylinder, shape border rotate=90, draw,minimum height=.9cm,minimum width=.9cm,left color=dbcolor!30,right color=dbcolor]
{~~~~};
\node (dblegend) at (.6,-.2) [anchor=west] {DB};

\node (action) at(1,1) [elem,left color=actioncolor!30, right color=actioncolor] {Actions};
\node (view) at(-1,1) [elem,left color=viewcolor!30, right color=viewcolor] {Queries};


\viewplace{viewplace}{(-1,2.1)}{[yshift=-.14cm,xshift=-.5cm]View places}

\node (place) at(0,2.1) [place] {};
\node (plegend) at (0,2.5) {Places};
\node (transition) at(1,2.1) [transition] {};
\node (plegend) at (1,2.5) [anchor=west,xshift=-.4cm] {Transitions};

\draw[->] (db) to node[anchor=east]{fetch} (view);
\draw[->] (action) to node[anchor=west]{update} (db);
\draw[->] (view) to node[anchor=east]{populate} (viewplace);
\draw[->] (transition) to node[anchor=west]{trigger} (action);
\draw[-stealth'] (2,2.3) to node[anchor=west,xshift=.2cm]{Arcs}  (2.5,2.3);
\draw[-] (2,2.05) to node[anchor=west,xshift=.2cm]{Read arcs} (2.5,2.05);
\draw[X-stealth'] (2,1.8) to node[anchor=west,xshift=.2cm]{Rollback arcs} (2.5,1.8);

  \begin{pgfonlayer}{background}
\node (data) at (-3,-.05) [back,fill=datacolor] {};
\node (activedata) at (-3,1) [back,fill=activedatacolor] {};
\node (netdata) at (-3,2.18) [back,fill=netcolor,minimum height=1.3cm] {};
\node (datalegend) at (-3,-.2) [anchor=east] {\bf persistence layer};
\node (datalegend2) at (3,-.2) [anchor=west] {\phantom{\bf persistence layer}};
\node (activedatalegend) at (-3,1) [anchor=east] {\bf data logic layer};
\node (netlegend) at (-3,2.18) [anchor=east] {\bf control layer};

 \end{pgfonlayer}

\end{tikzpicture}
}
\caption{The conceptual components of \dbnets\label{fig:dbnets}}
\end{figure}
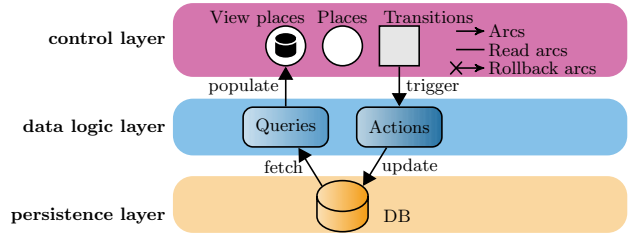

In our formal model, called \dbnet, we combine the distinctive features of CPNs and  relational databases into a coherent framework,  sketched in Figure~\ref{fig:dbnets}. The model is structured in three layers:
\begin{compactitem}[$\bullet$]
\item \emph{persistence layer}, capturing a full-fledged relational database with constraints, and used to store background data, and data that are persistent across cases.
\item \emph{control layer}, employing a variant of CPNs to capture the process control-flow, case data, and possibly the resources involved in the process execution.
\item \emph{data logic layer}, interconnecting in the persistence and the control layer.
\end{compactitem}
Thanks to the data logic, the control layer is supported in querying the underlying persistent data and tunes its own behavior depending on the obtained answers. Furthermore, the data logic may be exploited by the control layer to update the persistent data depending on the current state, the data locally carried by tokens, and additional data obtained from the external world.
We formalize the framework layer by layer, from the bottom to the top. 

\subsection{Persistence Layer}
The persistence layer maintains the relevant data in the domain of interest. To this end, we rely on standard relational databases equipped with constraints, in the spirit of \cite{BCDD13}. First-order (FO) constraints allow for the formalization of conventional database constraints, such as keys and functional dependencies, as well as semantic constraints reflecting the domain of interest. Differently from \cite{BCDD13}, though, we also consider data types, on the one hand resembling concrete logical schemas of relational databases (where table columns are typed), and on the other reconciling the persistence layer with the notion of ``color" in CPNs. 

\begin{notion}[Data type, type domain]\label{def:dtypes}
A \emph{data type} $\type$ is a pair $\tup{\dom_\type,\sigp_\type}$, where $\dom_\type$ is a \emph{value domain}, and $\sigp_\type$ is a finite set of \emph{predicate  symbols}. Each predicate symbol $S \in \sigp_\type$  comes with an arity $n_S$ and an $n$-ary predicate $S^\type \subseteq \dom_{\type}^n$ that rigidly defines its semantics. A \emph{type domain} is a finite set of data types.
\end{notion}
In the following, we use $\types$ to denote a type domain of interest, assuming that types in $\types$ are pairwise disjoint, that is, their domains do not intersect, and their predicate symbols are syntactically distinguished. This guarantees that a predicate symbol $S$ defined in some type of $\types$, is defined only in that type, which can be then unambiguously denoted, with slight abuse of notation, by $\ptype(S)$.   We also employ $\dom_{\types} = \bigcup_{\type \in \types} \dom_{\type}$.
Examples of data types are:
\begin{compactitem}[$\bullet$]
\item $\typename{string} : \tup{\strings,\set{=_s}}$, strings with the equality predicate;
\item $\typename{real} : \tup{\reals,\set{=_r,<_r}}$, real numbers with the usual comparison operators; 
\item $\typename{int} : \tup{\ints,\set{=_{int},<_{int},succ}}$, integers with the usual comparison operators, as well as the successor predicate. 
\end{compactitem}

We employ Definition~\ref{def:dtypes} to introduce a typed relational database. 
\begin{notion}[Typed relation schema, arity]
A \emph{$\types$-typed relation schema} is a pair $\tup{R,\vec{\type}}$, where $R$ is a relation name, and $\vec{\type}$ is a tuple of elements from $\types$, indicating the data types associated to each component of $R$. The number $|\vec{\type}|$ is the \emph{arity} of $R$. 
\end{notion}
\begin{notion}[Typed database schema]\label{def:typed-schema}
A \emph{$\types$-typed database schema} $\schema$ is a finite set of $\types$-typed relation schemas.
\end{notion}
For compactness, we represent a typed relation schema $\tup{R,\tup{\type_1,\ldots,\type_n}}$ using notation $R(\type_1,\ldots,\type_n)$. E.g., $\relname{Emp}(\typename{string},\typename{string},\typename{real})$ models a ternary relation for employees, where the first component is a string denoting the employee id, the second a string for her name, and the third a real for her salary.

\begin{notion}[Typed database instance, active domain]\label{def:typed-db}
Given a $\types$-typed database schema $\schema$, a \emph{$\types$-typed database instance $\I$ over $\schema$} is a \emph{finite} set of facts of the form $R(\cname{o_1},\ldots,\cname{o_n})$, such that 
\begin{inparaenum}[\it (i)]
\item $R(\type_1,\ldots,\type_n) \in \schema$, and 
\item for each $i \in \set{1,\ldots,n}$, we have $\cname{o_i}\in \dom_{\type_i}$.
\end{inparaenum}
Given a type $\type \in \types$, the \emph{$\type$-active domain of $\I$}, written $\adom[\type]{\I}$, is the set of values in $\dom_\type$ such that $\cname{o} \in \adom[\type]{\I}$ if and only if $\cname{o} \in \dom_\type$ and $\cname{o}$ occurs in $\I$. 
\end{notion}

%

We now turn to queries. As query language, we resort to standard first-order logic (FOL), interpreted under the active domain semantics \cite{Lib04}. This means that quantifiers are relativized to the active domain of the database instance of interest, guaranteeing that queries are domain-independent (actually, safe-range): their evaluation only depends on the values explicitly appearing in the database instance over which they are applied. Recall that this query language is equivalent to the well-known SQL standard \cite{AbHV95}. Since the relational structures we consider are typed, the logic is typed as well.

Given a type domain $\types$, we fix a countably infinite set $\varset$ of variables. Each variable is typed. To this end, we introduce a \emph{variable typing function} $\vartype: \varset \rightarrow \types$ mapping variables to their types. The typing function prescribes that $x$ may be substituted only by values taken from $\dom_{\vartype(x)}$. 
For compactness, the variable type may be explicitly shown using a colon notation $\typed{x}{\vartype(x)}$.

\begin{notion}[\fod query]
\label{def:fod-query}
	A \emph{(well-typed) \fod query} over a $\types$-typed database schema $\schema$ is a formula of the form:
\[
 Q ~::=~ S(\vec{y}) \mid R(\vec{z}) \mid  \lnot Q ~\mid~ Q_1\land Q_2 ~\mid~
              \exists x.Q, \text{ where}
\]
\begin{compactitem}[$\bullet$]
\item for $\vec{y} = \tup{y_1,\ldots,y_n}$, we have that $S/n$ is a predicate defined in $\sigp_\type$ for some $\type \in \types$, and for each $i \in \set{1,\ldots,n}$, we have that $y_i$ is either a value $\cname{o} \in \dom_{\type}$, or a variable $x \in \varset$ with $\vartype(x) = \type$;
\item for $\vec{z} = \tup{z_1,\ldots,z_m}$, we have that $R(\type_1,\ldots,\type_m)$  is a relation defined in $\schema$, and for each $i \in \set{1,\ldots,m}$, we have that $z_i$ is either a value $\cname{o} \in \dom_{\type_i}$, or a variable $x \in \varset$ with $\vartype(x) = \type_i$.
\end{compactitem} 
\end{notion}
We  use standard abbreviations $Q_1 \lor Q_2 = \neg (\neg Q_1 \land\neg Q_2)$, and $\forall x. Q = \neg \exists x. \neg Q$.

\begin{notion}[Free variable, boolean query]
A variable $x \in \varset$ is \emph{free} in a \fod query $Q$, if $x$ occurs in $Q$ but is not in the scope of any quantifier. We use $\free{Q}$ to denote the set of variables occurring free in $Q$. A \emph{boolean query} is a query without free variables.
\end{notion}
Given a query $Q$ such that $\free{Q} = \set{x_1,\ldots,x_n}$, we employ notation $\datq{\cname{Q_{name}}}{x_1,\ldots,x_n}{Q}$
 to emphasize the free variables of $Q$, and to fix a natural ordering over them. 
As usual, queries are used to extract answers from a database instance of interest.

\begin{notion}[Substitution]\label{def:substitution}
Given a set $X = \tup{x_1,\ldots,x_n}$ of typed variables, a \emph{substitution}  for $X$ is a function $\subst: X \rightarrow \dom_\types$ mapping variables from $X$ into values, such that for every $x \in X$, we have $\subst(x) \in \dom_{\vartype(x)}$. A \emph{substitution $\subst$ for a \fod query $Q$} is a substitution for the free variables of $Q$.
\end{notion} 
As customary, we may view a substitution $\subst$ for a query $Q$ simply as a tuple of values, assuming the natural ordering over the free variables of $Q$.
We denote by $Q\subst$ the boolean query obtained from $Q$ by replacing each free variable $x \in \free{Q}$ with the corresponding value $\subst(x)$. In the following, we apply substitutions to any structure containing variables.

\begin{notion}[Query entailment with active domain semantics]\label{def:query-semantics}
Given a $\types$-typed database schema $\schema$, a $\types$-typed instance $\I$ over $\schema$, a \fod query $Q$ over $\schema$, and a substitution $\subst$ for $Q$, we inductively define the relation  $\I$ \emph{entails $Q$ under $\subst$ with active domain semantics}, 
written $\qent{\I}{\subst}{Q}$, as:
 \[
 \begin{array}{l@{\textnormal{~if~~}}l}
 	\qent{\I}{\subst}{R(y_1,\ldots,y_n)} & R(y_1,\ldots,y_n)\subst \in \I\\
 	\qent{\I}{\subst}{S(y_1,\ldots,y_n)} & S(y_1,\ldots,y_n)\subst \in S^{\ptype(S)}\\
 	\qent{\I}{\subst}{\neg Q} & \qnent{\I}{\subst}{Q}\\
 	\qent{\I}{\subst}{Q_1 \land Q_2} & \qent{\I}{\subst}{Q_1} \text{ and } \qent{\I}{\subst}{Q_2}\\
 	\qent{\I}{\subst}{\exists x.Q} & \text{there exists } \cname{o} \in \adom[\vartype(x)]{\I} \text{ such that } \qent{\I}{\subst[x/\cname{o}]}{Q}\\ 	
 \end{array}
 \]
 where $\subst[x/\cname{o}]$ denotes the substitution obtained from $\subst$ by assigning $\cname{o}$ to $x$.\footnote{If $\subst(x)$ is defined, its value is replaced by $\cname{o}$, otherwise $\subst$ is extended so that $\subst(x) =\cname{o}$.}
\end{notion}

\begin{notion}[Query answers]
	Given a $\types$-typed database schema $\schema$, a $\types$-typed instance $\I$ over $\schema$, and a \fod query $Q(x_1,\ldots,x_n)$ over $\schema$, the set of \emph{answers to $Q$ in $\I$}, written $\ans(Q,\I)$, is the set of substitutions $\subst$ from the free variables of $Q$ to the active domain of $\I$, such that $Q$ holds in $\I$ under $\subst$:
	\[
	\ans(Q,\I) = 
	\left\{
	\left.
	\begin{array}{l}
		\subst: \set{x_1,\ldots,x_n} \rightarrow \\
		\qquad \adom[\vartype(x_1)]{\I} \times \ldots \times \adom[\vartype(x_n)]{\I}  \end{array}
	\right|
	\qent{\I}{\subst}{Q}
	\right\}
	\]
\end{notion}
When $Q$ is boolean, we write $\ans(Q,\I) \equiv \true$ if $\tup{} \in \ans(Q,\I)$,
or $\ans(Q,\I) \equiv \false$ if $\ans(Q,\I) = \emptyset$.
We are finally ready to define the persistence layer.

\begin{notion}[Persistence layer]\label{def:persistence-layer}
A $\types$-typed \emph{persistence layer} is a pair $\tup{\schema,\cset}$ where:
\begin{inparaenum}[\it (i)]
\item $\schema$ is a $\types$-typed database schema;
\item $\cset$ is a finite set $\{\Phi_1,...,\Phi_k\}$ of boolean \fod queries over $\schema$, modeling \emph{constraints over $\schema$}.
\end{inparaenum}
\end{notion}

The presence of constraints calls for a definition of which database instances are compliant by a given persistence layer, i.e., satisfy its constraints.
\begin{notion}[Compliant database instance]
Given a $\types$-typed persistence layer $\pers = \tup{\schema,\cset}$ and a $\types$-typed database instance $\I$, we say that $\I$ \emph{complies with} $\pers$ if:
\begin{inparaenum}[\it (i)]
\item $\I$ is defined over $\schema$;
\item $\I$ satisfies all constraints in $\cset$, that is, $\ans(\bigwedge_{\Phi \in \cset}\Phi,\I) \equiv \true$. 
\end{inparaenum} 	
\end{notion}

\begin{example}\label{ex:ticket-pl}
The persistence layer $\pers = \tup{\schema,\cset}$ is a fragment of an information system used by a company to handle the submission of tickets, and their management by employees.  $\schema$ employs types  $\typename{string}$ and $\typename{int}$ to define the following relation schemas:
\begin{compactitem}[$\bullet$]
\item $\relname{Emp}(\typename{string})$ lists employee (names); 
\item $\relname{Ticket}(\typename{int},\typename{string})$ models ticket (identifiers) and their description;
\item $\relname{Resp}(\typename{string},\typename{int})$ models which employees handle which tickets: $\relname{Resp}(\cname{e},\cname{1})$ indicates that the employee named $\cname{e}$ is responsible for ticket number $\cname{1}$.
\item $\relname{Log}(\typename{int},\typename{string},\typename{string})$ represents a log table storing information about all the tickets processed so far, also listing their responsible employees and their description.
\end{compactitem}
The persistence layer is also equipped with a set of constraints over $\schema$, expressing (primary) keys, foreign keys, functional dependencies, and multiplicity constraints. E.g., the ticket number provides the primary key for $\relname{Ticket}$, the second component of $\relname{Resp}$ references the primary key of $\relname{Ticket}$, and \emph{each employee can handle at most one ticket at a time}.
It is well-known that such constraints can be formalized in FO \cite{AbHV95}. E.g., the latter constraint may be formalized as: $\forall e,t_1,t_2. \relname{Resp}(e,t_1) \land \relname{Resp}(e,t_2) \rightarrow t_1 = t_2$.
\end{example}

\subsection{Data Logic Layer}\label{ex:ticket-dl}
The data logic layer provides a bidirectional ``interface'' to interact with a database instance complying with a persistence layer of interest. On the one hand, the data logic allows one to \emph{extract} data from the database instance using queries. On the other hand, it allows one to \emph{update} the database instance, adding and deleting possibly multiple facts at once, with a \emph{transactional} semantics: if the new database instance obtained after the update is still compliant with the persistence layer, the update is \emph{committed}, otherwise it is \emph{rolled back}. This approach is in line with how database management systems operate in practice.

To query the database instance, we use \fod queries as in Definition~\ref{def:fod-query}. To update the database instance, we instead resort to the literature on data-centric processes \cite{Vian09,CaDM13}, where \emph{actions} are typically used to apply CRUD (create-read-update-delete) operations over a relational database. Specifically, we adopt a minimalistic approach, keeping the actions as simple as possible. The approach is inspired by the well-known STRIPS language for planning, which has been adopted also in for data-centric processes \cite{AAA16}. More sophisticated forms of actions, as those in \cite{BCDD13}, can be seamlessly introduced. 

\begin{notion}[Action]
A \emph{(parameterized) action} over a $\types$-typed persistence layer $\tup{\schema,\cset}$ is tuple $\tup{\cname{n},\vec{p},F^+,F^-}$, where:
\begin{inparaenum}[\it (i)]
	\item $\cname{n}$ is the \emph{action name};
	\item $\vec{p}$ is a tuple of pairwise distinct typed variables from $\varset$, denoting the \emph{action (formal) parameters}.
	\item $F^+$ and $F^-$ respectively represent a finite set of $\schema$-facts over $\vec{p}$, to be \emph{added} to and \emph{deleted} from the current database instance.
\end{inparaenum}
	Given a typed relation $R(\type_1,\ldots,\type_n) \in \schema$, an $R$-fact over $\vec{p}$ 
	has the form $R(y_1,\ldots,y_n)$, such that for every $i \in \set{1,\ldots,n}$, $y_i$ is either a value $\cname{o} \in \dom_{\type_i}$, or a variable $x \in \vec{p}$ with $\vartype(x) = \type_i$. An $\schema$-fact is an $R$-fact for some relation $R$ from $\schema$.
\end{notion}
To access the different components of an action $\action = \tup{\cname{n},\vec{p},F^+,F^-}$, we use a dot notation: $\aname{\action} = \cname{n}$, $\apar{\action} = \vec{p}$, $\aadd{\action}= F^+$, and $\adel{\action}=F^-$.


We now turn to the semantics of actions. Actions are executed by grounding their parameters to values. Given an action $\action$ and a (parameter) substitution $\subst$ for $\action$, we call \emph{action instance $\action\subst$} the (ground) action resulting from $\action$ by substituting its parameters with corresponding values, as specified by $\subst$.

\begin{notion}[Action instance application]
Let $\pers = \tup{\schema,\cset}$ be a $\types$-typed persistence layer, $\I$ be a $\types$-typed database instance $\I$ compliant with $\types$, $\alpha$ be an action over $\pers$, and $\subst$ be a substitution for $\apar{action}$. The \emph{application} of $\action\subst$ on $\I$, written $\doact{\action\subst}{\I}$, is a database instance over $\schema$ obtained as $(\I\setminus F^-_{\action\subst})\cup F^+_{\action\subst}$, where:
\begin{inparaenum}[\it (i)]
\item $F^-_{\action\subst} = \bigcup_{R(\vec{y}) \in \adel{\action}} R(\vec{y})\subst$;
\item $F^+_{\action\subst} = \bigcup_{R(\vec{y}) \in \aadd{\action}} R(\vec{y})\subst$.
\end{inparaenum}
We say that $\action\subst$ can be \emph{successfully applied} to $\I$ if $\doact{\action\subst}{\I}$ complies with $\pers$.
 \end{notion}
 The application of an action instance amounts to ground all the facts contained in the definition of the action as specified by the given substitution, then applying the update on the given database instance, giving priority to additions over deletions (this is a standard approach, which unambiguously handles the situation in which the same fact is asserted to be added and deleted).

The data logic simply exposes a set of queries and a set of actions that can be used by the control layer to obtain data from the persistence layer, and to induce updates on the persistence layer. 

\begin{notion}[Data logic layer]\label{def:data-logic-layer}
	Given a $\types$-typed persistence layer $\pers$, a \emph{$\types$-typed data logic layer over $\pers$} is a pair $\tup{\queries,\actions}$, where:
	\begin{inparaenum}[\it (i)]
	\item $\queries$ is a finite set of \fod queries over $\pers$;
	\item $\actions$ is a finite set of actions 	over $\pers$.
	\end{inparaenum}
\end{notion}

\begin{example}
We make the scenario of Example~\ref{ex:ticket-pl} operational, introducing a data logic layer $\dl$ over $\pers$. $\dl$ exposes two queries to inspect the persistence layer:
\begin{compactitem}[$\bullet$]
\item $\datq{\cname{Q_{e}}}{e}{\relname{Emp}(e) \land \neg \exists t.\relname{Resp}(e,t)}$, to extract \emph{idle} employees;
\item $\datq{\cname{Q_{t}}}{t,d}{\relname{Ticket}(t,d)}$, to extract tickets and their description.
\end{compactitem}
In addition, $\dl$ provides three main functionalities to manipulate tickets in persistence layer: ticket registration, assignment/release, and logging. Such functionalities are realized through four actions (where, for simplicity, we blur the distinction between an action and its name).
The registration of a new ticket is managed by an action $\actname{reg}$ that, given an integer $t$, and two strings $e$ and $d$, ($\apar{\actname{reg}}=\tup{\pname{t,e,d}}$, simultaneously creates a ticket identified by $\pname{t}$ and described by $\pname{d}$ into the persistence layer, and assigns the employee identified by $\pname{e}$ to such ticket (thus making her \emph{busy}):
\[
\adel{\actname{reg}}=\set{\relname{Emp}(\pname{e},\cname{idle})} 
\qquad 
\aadd{\actname{reg}}=\set{\relname{Ticket}(\pname{t,d}),\relname{Resp}(\pname{e},\pname{t})}
\]
Two specular actions $\actname{assign}$ and $\actname{release}$ are exposed to assign or release a ticket to/from an employee, making her busy or idle. Both actions take as input a string for the employee name and an integer for a ticket it ($\apar{\actname{assign}}= \apar{\actname{release}}=\tup{\pname{e,t}}$), and update $\pname{e}$ by removing or adding that $e$ is responsible of $t$:
\[
\adel{\actname{release}}=\aadd{\actname{assign}}=\set{\relname{Resp}(\pname{e,t})}
\qquad
\aadd{\actname{release}}=\adel{\actname{assign}}=\emptyset
\]
Finally, an action $\actname{log}$ with $\apar{\actname{log}}=\tup{\pname{t},\pname{e},\pname{d}}$ is exposed to flush the information related to a ticket into a log table. The action erases all information about the ticket, and logs that it has been processed, also recalling its employee and description:
\[
\adel{\actname{log}}=\set{\relname{Ticket}(\pname{t,d}),\relname{Resp}(\pname{e,t})}
\qquad
\aadd{\actname{log}}=\set{\relname{Log}(\pname{t,e,d})}
\]
\end{example}

\subsection{Control Layer}
The control layer employs a variant of CPNs to capture the process control flow, and how it interacts with an underlying persistence layer through the functionalities provided by the idata logic. The spirit is to conceptually ground CPNs by adopting a data-oriented approach. This is done by  introducing dedicated constructs exploiting such functionalities, as well as simple, declarative patterns to capture the typical token consumption/creation mechanism of CPNs. 

Before introducing the different constitutive elements of the control layer together with their graphical appearance, we fix some preliminary {notion}s. 
We consider the standard notion of a \emph{multiset}. Given a set $A$,  the \emph{set of multisets} over $A$, written $\mult{A}$, is the set of mappings of the form $m:A\rightarrow \mathbb{N}$.
Given a multiset $S \in \mult{A}$ and an element $a \in A$, $S(a) \in \mathbb{N}$ denotes the number of times $a$ appears in $S$.  Given $a \in A$ and $n \in \mathbb{N}$, we write $a^n \in S$ if $S(a) = n$.  We also consider the usual operations on multisets. Given $S_1,S_2 \in \mult{A}$:
\begin{inparaenum}[\it (i)]
\item $S_1 \subseteq S_2$ (resp., $S_1 \subset S_2$) if $S_1(a) \leq S_2(a)$ (resp., $S_1(a) < S_2(a)$) for each $a \in A$;
\item $S_1 + S_2 = \set{a^n \mid a \in A \text{ and } n = S_1(a) + S_2(a)}$;
\item if $S_1 \subseteq S_2$, $S_2 - S_1 = \set{a^n \mid a \in A \text{ and } n = S_2(a) - S_1(a)}$;
\item given a number $k \in \mathbb{N}$, $k \cdot S_1 = \set{a^{kn} \mid a^n \in S_1}$.\footnote{Hence, given a multiset $S$, we have $0 \cdot S = \emptyset$.}
\end{inparaenum}

\smallskip
\noindent
\textbf{Places.}
The control layer contains a finite set $\places$ of places, which in turn are classified in two groups. On the one hand, so-called  
\emph{control places} play the role of standard places in classical Petri nets: they represent conditions/states of a dynamic system. On the other hand, so-called \emph{view places} are used as an interface to the underlying persistence layer, so as to make the persistent data available to the control layer. We then have $\places = \cplaces \uplus \vplaces$, where $\cplaces$ and $\vplaces$ respectively denote the set of control and view places.
Graphically, we depict control places using the standard notation:
\resizebox{.35cm}{!}{\tikz{\node [circle,draw,thick] at (0,0) {};}}. 
We instead decorate view places as: 
\resizebox{.35cm}{!}{\tikz{\node [circle,draw,very thick,minimum width=.6cm, minimum height=.6cm] at (0,0) {}; \dbicon{(0,0)}{.4cm}{.3cm}}}.

In the spirit of CPNs, the control layer assigns to each place a color, which in turn combines one or more data types from a type domain $\types$. Formally, a \emph{$\types$-color} is a cartesian product $\type_1 \times \ldots \times \type_m$, where for each $i \in \set{1,\ldots,m}$, we have $\type_i \in \types$. We denote by $\colors$ the set of all possible $\types$-colors.

\begin{notion}[Color assignment]\label{def:coloring}
A \emph{$\types$-color assignment} over places $\places$ is a function $\coloring: \places \rightarrow \colors$ mapping each place $p \in \places$ to a corresponding $\types$-color. 
\end{notion}

As for control places, it is well-known that the coloring mechanism can be exploited to realize a plethora of conceptual abstractions on top of the control flow. We mention here the two most important abstractions in our setting: 
\begin{inparaenum}[\it (i)]
\item cases and their data, and 
\item resource.
\end{inparaenum}
 A \emph{case} represents a specific process instance, and its \emph{case data} \cite{RHE05} are local data whose scope is the case itself, and that are used to store important information for the progression of the case. Such data may be either extracted from the underlying persistence layer, or obtained by interacting with the external environment (e.g., human users, external services, or data generators).
\emph{Resources} represent actors able to handle the execution of tasks. They are also typically associated to data attributes (e.g., id, role, group). Tasks typically consume (certain kinds of) resources when executed, and this implicitly affect the degree of concurrency in the progression of cases, as well as the possibility of spawning new cases.

The fact that control places are colored implies that whenever a token is assigned to a control place, it must carry a data tuple whose types match component-wise the place color.
It is worth noting that a colored place may be interchangeably considered as a specific state/condition within the control layer, or as a special relation schema used to enrich the persistence layer with control-related information. Similarly, a token distributed over a place may be interchangeably seen as a thread of control located in that state, or as a tuple assigned to the relation schema represented by that place. 

As discussed above, control places host tokens carrying local data. Obviously, the control layer also requires to query persistent data, using them to decide how to route tokens when it comes to business decisions, or to assign them to case data. We want to support both possibilities, but clearly separating the data retrieved from the persistence layer, from those carried by tokens. This is why we distinguish view places from control places. Each view place exposes to the control layer a portion of the data stored in the persistence layer. Formally, this is done by equipping the view place with a query defined in the data logic layer. 

\begin{notion}[Query assignment]\label{def:query-assignment}
Given a data logic layer $\dl = \tup{\queries,\actions}$, a \emph{query assignment} from view places $\vplaces$ to queries $\queries$ is a function $\quass: \vplaces \rightarrow \queries$ mapping each view place $p \in \vplaces$ with $\coloring(p) = \type_1\times\ldots\times\type_n$ to a query $Q(x_1,\ldots,x_n)$ from $\queries$, such that the color of $p$ component-wise matches with the types of the free variables in $Q$: for each $i \in \set{1,\ldots,n}$, we have $\type_i = \vartype(x_i)$.
\end{notion}
A view place may be seen as a normal place, whose color is implicitly obtained by the types of the free variables of the query, considered with their natural ordering. However, tokens are not arbitrarily attached to it: at a given time, the tokens it contains represent the answers to the query it is associated to. All such tokens are only ``virtually" present in the control layer, and in fact they cannot be consumed within the control layer itself, but only accessed in a read-only way. Notice, however, that the content of the view place is not immutable: it changes whenever the data it fetches from the persistence layer are updated.

\smallskip
\noindent
\textbf{Transitions.}
As customary, in our model transitions represent atomic units of work within the control layer, thus providing the fundamental building block to describe the dynamics of a process. As usual, they are depicted using a square notation: \resizebox{.35cm}{!}{\tikz{\node [rectangle,draw,thick] at (0,0) {};}}.
In our setting, they simultaneously account for three different aspects: the token consumption/production mechanism of CPNs, the injection of possibly fresh data from the external environment a l\`a $\nu$-Petri nets \cite{RVFE11}, and the impact on the underlying persistence layer. 

We start with the consumption of tokens. This is modeled through input arcs connecting places to transitions, together with inscriptions that declaratively match tokens and their data. To this end, we build on the approach adopted in variants of data nets \cite{RVFE11,Las16,AAA16}: an inscription is just a multiset of tuples over a given set of typed variables. Each tuple matches with a token present in the input place, and the variables therein are bound, component-wise, to the data carried by such a token. In addition, if the input place is a control place, the token is consumed upon firing, whereas if the place is a view place, it is only inspected. 
Graphically, we adopt the following conventions. An input arc from a control place is depicted as usual:
\resizebox{1.2cm}{!}{\tikz{\node (s) [circle,draw,thick] at (0,0) {};\node (t) [rectangle,draw,thick] at (1,0) {};\draw[-stealth',thick] (s) -- (t);}}.
An input arc from a view place is instead depicted using the read-arc notation: \resizebox{1.2cm}{!}{\tikz{\node (s) [circle,draw,very thick,minimum width=.6cm, minimum height=.6cm] at (0,0) {}; \dbicon{(0,0)}{.4cm}{.3cm}\node (t) [rectangle,draw,very thick,minimum width=.5cm,minimum height=.5cm] at (1.5,0) {};\draw[-,very thick] (s) -- (t);}}.

The overall consumption/inspection of  tokens and the data they carry along all arcs incoming into a transition constitutes a \emph{firing mode} for that transition.
In the context of a transition definition, we call a tuple of typed variables (as well as, possibly, values) \emph{inscription}. We denote the set of all possible inscriptions over set $\mathcal{Y}$ as $\tuples{\mathcal{Y}}$. In addition, we denote the set of variables appearing inside an inscription $\omega \in \tuples{\mathcal{Y}}$ as $\varsin{\omega}$, and we extend such notation to sets and multisets of inscriptions.


\begin{notion}[Input flow]\label{def:input-flow}
An \emph{input flow} from places $\places$ to transitions $\transitions$ is a function $\inflow: \places \times \transitions \rightarrow \mult{\tuples{\varset}}$ assigning multisets of inscriptions (over variables $\varset$) to input arcs, such that all such inscriptions are compatible with their input places. An inscription $\tup{x_1,\ldots,x_m}$ is \emph{compatible} with a place $p$ if $\coloring(p) = \type_1 \times \ldots\times \type_m$, such that for every $i \in \set{1,\ldots,m}$, we have $\vartype(x_i) = \type_i$.
%
%
%
\end{notion}
%
%
%
%
%
%
Graphically, we do not depict input arcs whose inscription is $\emptyset$.
We  define the \emph{input variables} of $t$, written $\invars{t}$ as the set of all variables occurring on input arc inscriptions for $t$:\begin{center}
	$\invars{t} = \set{x \in \varset \mid \text{there exists }p \in \places \text{ such that }x \in \varsin{\inflow(\tup{p,t})}}$.
\end{center}

The set $\invars{t}$ gives an indication about which input data elements are accessed when a transition fires. The multiple usage of the same variable in an inscription, or in 
inscriptions attached to different arcs incident to a transition, captures the requirement of \emph{matching} the same data object in different tokens, allowing the transition to fire only if the accessed tokens carry the \emph{same} data value. This mirrors the notion of join used when querying relational data. 
In general, though, the modeler may require to specify additional constraints over such input data to allow firing the transition. To this end, we introduce guards.

\begin{notion}[Guard]
A $\types$-typed \emph{guard} is a formula of the form:
\[
\varphi~::=~\true \mid S(\vec{y}) \mid \neg \varphi  \mid \varphi_1 \land \varphi_2
\]
where, for $\vec{y} = \tup{y_1,\ldots,y_n} \subseteq \varset$, we have that $S/n$ is a predicate defined in $\sigp_\type$ for some $\type \in \types$, and for each $i \in \set{1,\ldots,n}$, we have that $y_i$ is either a value $\cname{o} \in \dom_{\type}$, or a variable $x_i \in \varset$ with $\vartype(x_i) = \type$. 
\end{notion}
We denote by $\guards{\types}$ the set of all possible $\types$-typed guards. Additionally, with a slight abuse of notation, given guard $\varphi$ we denote by $\varsin{\varphi}$ the set of variables occurring in $\varphi$.
Guards may be seen as the quantifier- and relation-free fragment of \fod queries (cf.~Definition~\ref{def:fod-query}). Consequently, their semantics is inherited from Definition~\ref{def:query-semantics} (considering the empty database instance). 
Guards are attached to transitions, and defined over their input variables, thus being an additional filter on the data that can be matched to the input inscriptions.
\begin{notion}[Transition guard assignment]\label{def:guard}
A $\types$-typed \emph{transition guard assignment} over transitions $\transitions$  is a function $\guass:  \transitions \rightarrow \guards{\types}$ assigning to each   transition $t \in T$ a  $\types$-typed guard $\varphi$, such that $\varsin{\varphi} \subseteq \invars{t}$.
\end{notion}

We now concentrate on the effect of firing a transition, which may simultaneously impact the control layer and the underlying persistence layer. Such an effect is tuned by the input variables attached to the transition, as well as additional data obtained from the external environment. Injection of external data is crucial for two reasons \cite{CaDM13,MonC16,AAA16}. First, during the execution of a case, input data may be dynamically acquired from human users or external services, and used later on; this is, e.g., what happens when a user form needs to be filled before continuing with the case execution, then deciding how to route the case depending on the inserted data. Second, fresh identifiers may be injected into the system, e.g., to explicitly distinguish tokens via certain data attributes, or to insert a new tuple in the underlying database instance (which typically requires to create a distinctive primary key for that tuple). We call these two types of external inputs \emph{arbitrary external inputs} and \emph{fresh external inputs}. To account for arbitrary external inputs in the context of a transition, we just employ ``normal" variables distinct from those used in the input inscriptions. To account for fresh external inputs, we employ the well-known mechanism adopted in $\nu$-Petri nets \cite{RVFE11,MonR16}. In particular, we introduce a countably infinite set $\nuvarset$ of $\types$-typed \emph{fresh variables}. To guarantee an unlimited provisioning of fresh values, we impose that for every variable $\nu \in \nuvarset$, we have that $\dom_{\vartype(\nu)}$ is countably infinite.

From now on, we fix a countably infinite set of $\types$-typed variable $\vars$, obtained as the disjoint union of ``normal" variables $\varset$ and fresh variables $\nuvarset$. In formulae,   $\vars = \varset \uplus \nuvarset$.
Let us first focus on the impact of transition firing on the underlying persistence layer. This is, again, mediated by the data logic, exploiting in particular the actions it exposes. Specifically, a transition can bind to an action, using variables from $\vars$ as ``actual" parameters. In this light, data passing from the control to the persistence layer is captured by re-using the same variable inside an input inscription and an action binding for the same transition. When the  transition  fires,  actual parameters are substituted with concrete data values, instanating the action and allowing for its further invocation.

\begin{notion}[Action assignment]\label{def:action-assignment}
Given a data logic layer $\dl = \tup{\queries,\actions}$, 
	an \emph{action assignment} from transitions $\transitions$ to actions $\actions$ is a partial function $\aass: \transitions \rightarrow \actions \times \tuples{\vars \cup \dom_\types}$, where $\aass(t)$ maps $t$ to an action $\action \in \actions$ together with a (binding) inscription compatible with $\action$. An inscription $\tup{y_1,\ldots,y_m}$ is compatible with $\action$ if $\apar{\action} = \tup{z_1,\ldots,z_m}$ and, for each $i \in \set{1,\ldots,m}$, we have $\vartype(y_i) = \vartype(z_i)$ if $y_i$ is a variable from $\vars$, or $y_i \in \dom_{\vartype(z_i)}$ if $y_i$ is a value from $\dom_\types$. 
\end{notion}
The action assignment provides a distinctive feature of our model, namely the ability of the  control layer to invoke an action applied to the underlying persistence layer. This, however, does not in general guarantee that the action invocation will actually turn into an update over the persistence layer. Recall in fact that an action instance is applied transactionally: if it produces a new database instance that is compliant with the persistence layer, the action instance \emph{succeeds} and the update is committed; if, instead, some constraints is violated, the action instance \emph{fails} and the update does not take place.

Lastly, we consider the effect of transitions on the control layer itself, defining which tokens have to produced, together with the data they will carry, and to which places such tokens have to be assigned. This is done by mirroring the definition of input flow (cf.~Definition~\ref{def:input-flow}), with two distinctions. First, output arcs connect transitions to control places only, as view places cannot be explicitly modified within the control layer. Second, the inscriptions attached to output arcs may mention not only input variables, but also:
\begin{inparaenum}[\it (i)]
\item values, allowing for constructing tokens that carry explicitly specified data;
\item fresh variables, allowing for constructing tokens that carry data not already present in the net, nor in the underlying database instance.	
\end{inparaenum}

\begin{notion}[Output flow]\label{def:output-flow}
An \emph{output flow} from transitions $\transitions$ to control places $\cplaces$ is a  function $\outflow: \transitions \times \cplaces \rightarrow \mult{\tuples{\vars \cup \dom_\types}}$ assigning multisets of inscriptions to output arcs, such that all such inscriptions are compatible with their output places (as defined in Definition~\ref{def:input-flow}).
\end{notion}
We do not depict output arcs graphically when their inscription is $\emptyset$.
We  define the \emph{output variables} of $t$, written $\outvars{t}$, as the set of variables occurring in the action assignment for $t$ (if any), and in its output arc inscriptions:
  \[\outvars{t} = \begin{array}[t]{ll}
  	& \set{x \in \vars \mid \aass(t) \text{ is defined as } \tup{\alpha,\omega}, \text{ and } x \in \varsin{\omega}}\\  
  	\cup &
 	\set{x \in \vars \mid \text{there exists }p \in \places \text{ such that }x \in \varsin{\outflow(\tup{t,p})}}.
  	\end{array}  
    \]
With this notion at hand, we can obtain the \emph{external variables} of transition $t$ as $\outvars{t} \setminus \invars{t}$. Each such variable $x$ is not bound by any input inscription, and can consequently be assigned arbitrarily (if $x \in \varset$), or to a fresh value (if $x \in \nuvarset$). Among such variables, we explicitly refer to the fresh variables attached to $t$, using notation $\freshvars{t}$. Mathematically, $\freshvars{t} = \outvars{t} \cap \nuvarset$.
As discussed before, firing a transition may incur in the instantiation and invocation of an action from the data logic layer, and the so-obtained action instance may or not result in an actual update. To raise awareness of the control layer about these two radically different outcomes, we introduce two separate output flows: a normal output flow, capturing the actual effect of a transition on the control flow when its attached action succeeds, and a \emph{rollback flow}, capturing the actual effect of a transition on the control flow when its attached action fails. With this distinction, the control layer can fine-tune its own behavior in accordance with the transactional semantics of the persistence layer, e.g., taking a standard or a compensation route depending on the outcome of the action.
To graphically distinguish \emph{normal output arcs} from \emph{rollback output arcs}, we proceed as follows. We depict the former as usual: 
\resizebox{1.2cm}{!}{\tikz{\node (s) [rectangle,draw,thick] at (0,0) {};\node (t) [circle,draw,thick] at (1,0) {};\draw[-stealth',thick] (s) -- (t);}}. 
Instead, we decorate the latter with an ``x":
\resizebox{1.2cm}{!}{\tikz{\node (s) [rectangle,draw,thick] at (0,0) {};\node (t) [circle,draw,thick] at (1,0) {};\draw[x-stealth',thick] (s) -- (t);}}.

We are finally in the position of defining the control layer.

\begin{notion}[Control layer]\label{def:control-layer}
A $\types$-typed \emph{control layer} over a data logic layer $\dl = \tup{\queries,\actions}$ is a tuple
$\tup{\places,\transitions,\inflow,\outflow,\rbflow,\coloring,\quass,\guass,\aass}$, where:
\begin{compactitem}[$\bullet$]
\item $\places = \cplaces \uplus \vplaces$ is a finite set of control places constituted by control places $\cplaces$ and view places $\vplaces$;
\item $\transitions$ is a finite set of transitions, such that $\transitions \cap \places= \emptyset$;
\item $\inflow$ is an input flow from $\places$ to $\transitions$ (cf.~Definition~\ref{def:input-flow});
\item $\outflow$ and $\rbflow$ are two output flows from $\transitions$ to $\cplaces$ (cf.~Definition~\ref{def:output-flow}), respectively called \emph{normal output flow}and \emph{rollback flow};
\item $\coloring$ is a color assignment over $\places$  (cf.~Definition~\ref{def:coloring});
\item $\quass$ is a query assignment from $\vplaces$ to $\queries$ (cf.~Definition~\ref{def:query-assignment});
\item $\guass$ is a transition guard assignment over $\transitions$ (cf.~Definition~\ref{def:guard});
\item $\aass$ is an action assignment from $\transitions$ to $\actions$ (cf.~Definition~\ref{def:action-assignment}).
\end{compactitem}
\end{notion}

\subsection{DB-nets}
We now put the three layers together, providing a formal definition for \dbnets.
\begin{notion}[Db-net]\label{def:db-net}
A \emph{db-net} is a tuple $\tup{\types,\pers,\dl,\cl}$, where:
\begin{compactitem}[$\bullet$]
\item $\types$ is a type domain	(cf.~Definition~\ref{def:dtypes});
\item $\pers$ is a $\types$-typed persistence layer	(cf.~Definition~\ref{def:persistence-layer});
\item $\dl$ is a $\types$-typed data logic layer over $\pers$ 	(cf.~Definition~\ref{def:data-logic-layer});
\item $\cl$ is a $\types$-typed control layer over $\dl$ 	(cf.~Definition~\ref{def:control-layer}).
\end{compactitem}

\end{notion}
We close the section by equipping our running example with a control layer.

\begin{example}\label{ex:order-management}
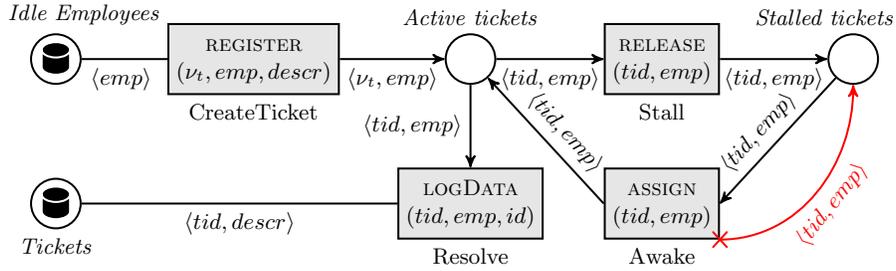
\begin{figure}[t!]
\centering
\resizebox{\hsize}{!}{
\begin{tikzpicture}[->,>=stealth',auto,x=1.8cm,y=1.7 cm,thick]
  \viewplace{employees}{(-1.75,0)}{above,xshift=.4cm:\emph{Idle Employees}};
  \node[transition, 
  label=below:CreateTicket] (create) at (-.2,0) {$\begin{array}{@{}c@{}}
                 \text{\actname{register}}\\
  				(\nu_{t},emp,descr)
               \end{array}
               $ };

\node[place,label={above:\emph{Active tickets}}] (active) at (1.5,0) {};
\node[transition,label=below:Stall] (stall) at (3,0) {$\begin{array}{@{}c@{}}
                 \text{\actname{release}}\\
  				(tid,emp)
               \end{array}
               $ };
\node[transition,label=below:Awake] (awake) at (3,-1.2) {$\begin{array}{@{}c@{}}
                 \text{\actname{assign}}\\
  				(tid,emp)
               \end{array}
               $ };
\node[place,label={above,xshift=-.4cm:\emph{Stalled tickets}}] (stalled) at (4.5,0) {};
\node[transition,label=below:Resolve] (log) at (1.5,-1.2) {$\begin{array}{@{}c@{}}
                 \text{\actname{logData}}\\
  				(tid,emp,id)
               \end{array}
               $ };
\viewplace{tickets}{(-1.75,-1.2)}{below:\emph{Tickets}};

\path
(create) edge node[below] {$\tup{\nu_{t},emp}$}  (active)
(active) edge[] node[below] {$\tup{tid,emp}$}(stall)
(stall) edge[] node[below] {$\tup{tid,emp}$}(stalled)
(stalled) edge node[sloped,above,xshift=-1mm] {$\tup{tid,emp}$}(awake.east)
(awake.west) edge node[sloped,above,xshift=1mm] {$\tup{tid,emp}$}(active)
(active) edge node[left] {$\tup{tid,emp}$} (log)

;

\path[-]
(employees) edge node[below,xshift=-.5mm] {$\tup{emp}$} (create)
(log) edge node[below] {$\tup{tid,descr}$} (tickets)

;
\path[X-stealth',thick,color=red]
(awake.south east) edge[out=0,in=-90] node[below,sloped] {$\tup{tid,emp}$}(stalled)
;
\end{tikzpicture}
}
\caption{The control layer of a \dbnet for ticket management. 
In CreateTicket, $\nu_t$ is a fresh input variable, and $descr$ is an arbitrary input variable.} \label{fig:wf-data-net}
\end{figure}

Figure~\ref{fig:wf-data-net} shows the control layer of a \dbnet $\dbn$, using the persistence layer $\pers$ defined in Example~\ref{ex:ticket-pl} and the data logic layer $\dl$ defined in Example~\ref{ex:ticket-dl}. The control layer realizes a simple ticket processing workflow, where tickets are created, manipulated, and finally resolved. In spite of its simplicity, $\dbn$ already shows many distinctive features of our model. We intuitively describe the control layer moving from left to right and from top to bottom. Each case of this process is constituted by a ticket and its responsible employee. A ticket is created by the CreateTicket transition, which requires the presence of an idle employee to be fired. Since this condition needs to inspect the persistence layer so as to retrieve idle employees, we model it through a view place associated to query $\cname{Q_{e}}$ from $\dl$. Notice that if no employee is currently idle, then CreateTicket is not enabled. Upon firing CreateTicket for a given idle employee, a fresh ticket identifier is generated using fresh variable $\nu_t$, and a ticket description is obtained through the ``external" input variable $descr$. All such data are bound to action $\actname{register}$, which is applied when the transition fires. Among the effects of $\actname{register}$, there is one asserting that the selected employee becomes responsible for the newly created ticket. This indirectly implies that such an employee is not present anymore in the view place for idle employees. The ticket id, together with its responsible employee, represent the case and its data. The two control places \emph{Active Tickets} and \emph{Stalled Tickets} have color $\typename{int} \times \typename{string}$, and model two distinct states in which tickets may be. Such states are important only within the evolution of cases, and are therefore not propagated to the underlying persistence layer. An active ticket may be ``stalled" if the employee is currently unable to resolve it. Executing the stall transition has a twofold effect. Within the control layer, the ticket is moved from active to stalled. Within the persistence layer, its responsible employee is released. Interestingly, the relation of responsibility is now only recalled within the control layer. A stalled ticket may be revived, by inserting such a relation back into the persistence layer. This is captured by the Awake transition, which mirrors the effect of the Stall transition. However, there is a particularly interesting aspect here. When a ticket $t_1$ is stalled, its responsible employee $e$ is released and becomes idle. She may be then selected as responsible of a newly created ticket $t_2$. Due to the constraints present in $\pers$, the indirect effect of this situation is that $t_1$ cannot be awaken unless $t_2$ is either stalled or resolved. In fact, awakening $t_1$ in a situation where $t_2$ is active would violate the requirement that $e$ is responsible of at most one ticket. For this reason, we enrich the Awake transition with a rollback output arc, which brings back the ticket to the stalled state if it is awaken in the ``wrong" moment.  For example, if $t_1$ is awaken while $t_2$ is active, the application of $\actname{assign}$ applied to $\tup{t_1,e}$ will fail, consequently bringing $t_1$ back to stalled.
Finally, an active ticket may be resolved. This has a twofold effect. On the one hand, the token carrying the ticket and its responsible employee is removed from the net. On the other hand, the case information is logged into the persistence layer. However, logging also requires to retrieve the description of the ticket. To this end, we employ a second view place accessing tickets and their description by exploiting $\cname{Q_{t}}$ from $\dl$. By using the same variable $tid$ in the two input inscriptions of the Resolve transition, we realize a join, thus inspecting the view place and extracting the description of $tid$.
%

\end{example}
\section{Execution Semantics}
The execution semantics of a \dbnet simultaneously accounts for the progression of a database instance compliant with the persistence layer of the net, and for the evolution of a marking over the control layer of the net. Such two information sources affect each other via the data logic layer: the database instance exposes its own data through view places, influencing the current marking and the enabled transitions; the marking over the control layer determines which transitions may be fired, in turn triggering updates the database instance. 
We start by formalizing the notion of marking over the control layer of the \dbnet. A marking distributes tokens over the places of the net, so that each token carries data that are compatible with the color of the place in which that token resides. In this light, tokens are nothing else than tuples of values over the place colors. In addition, the marking of a view place must  correspond to the answers obtained by issuing its associated query over the underlying database instance. 

\begin{notion}[Marking]
A \emph{marking} of a $\types$-typed control layer  $\cl = \tup{\places,\transitions,\inflow,\outflow,\rbflow,\coloring,\quass,\guass,\aass}$ is a function $\marking: \places \rightarrow \mult{\tuples{\types}}$ mapping each place $p \in \places$ to a corresponding multiset of $p$-compatible tuples using data values from $\types$. A tuple $\tup{\cname{o_1},\ldots,\cname{o_n}}$ is $p$-compatible if $\coloring(p)$ is of the form $\tup{\type_1,\ldots,\type_n}$, and for every $i \in \set{1,\ldots,n}$, we have $\cname{o_i} \in \dom_{\type_i}$. Given a database instance $\I$, we say that $\marking$ is \emph{aligned} to $\I$ via $\quass$ if the tuples it assigns to view places exactly correspond to the answers of their corresponding queries over $\I$: for every view place $v \in \places$ and every $v$-compatible tuple $\vec{\cname{o}}$, we have that $\vec{\cname{o}} \in \marking(v)$ if and only if $\vec{\cname{o}} \in \ans(\quass(v),\I)$.
\end{notion}

We mirror the notion of active domain as provided in Definition~\ref{def:typed-db} to the case of markings. Given a type $\type \in \types$, the \emph{$\type$-active domain} of a marking $m$, written $\adom[\type]{m}$, is the set of values in $\dom_\type$ such that $\cname{o} \in \adom[\type]{m}$ if and only if there exists $p$ such that $\cname{o}$ occurs in $m(p)$.
From the practical point of view, one may consider the marking of control places to be initially defined by the modeler, and then evolved by the control layer, while the marking of view places computed on-the-fly from the underlying database instance when needed.

In \dbnets, then, both the persistence layer and the control layer are stateful: during the execution, the persistence layer is associated to a database instance, while the control layer to a marking aligned with that database instance. 

\begin{notion}[Snapshot]
	Given a \dbnet $\dbn = \tup{\types,\pers,\dl,\cl}$ with control layer $\cl = \tup{\places,\transitions,\inflow,\outflow,\rbflow,\coloring,\quass,\guass,\aass}$, a \emph{snapshot} of $\dbn$ (also called \emph{$\dbn$-snapshot}) is a pair $\dbstate{\I}{\marking}$, where $\I$ is a database instance compliant with $\pers$, and $\marking$ is a marking of $\cl$ aligned to $\I$ via $\quass$.
\end{notion}

As customary for CPNs, the firing of a transition $t$ in a snapshot is defined w.r.t.~a so-called \emph{binding} for $t$, that is, a substitution $\mode: \varsin{t} \rightarrow \dom_{\types}$, where $\varsin{t} = \invars{t} \cup \outvars{t}$. However, to properly enable the firing of $t$, the binding $\sigma$ must guarantee a number of properties:
\begin{compactenum}
\item agreement with the distribution of tokens over the places, in accordance with the inscriptions on the corresponding input arcs;
\item satisfaction of the guard attached to $t$;
\item proper treatment of fresh variables, guaranteeing that they are substituted with values that are pairwise distinct, and also distinct from all the values present in the current marking, as well as in the current database instance.
\end{compactenum}
To formalize these conditions, we need the preliminary notion of inscription binding. 
Given an inscription (i.e., multiset of tuples of variables) $\omega \in \mult{\tuples{\vars \cup \dom_\types}}$, and a substitution $\theta$ defined over a set $X$ of variables containing all variables occuring in $\omega$, the \emph{inscription binding} of $\omega$ under $\subst$ is a multiset $\ibind{\subst}{\omega}$ from $\mult{\tuples{\types}}$ defined as follows: $\tup{\cname{o_1},\ldots,\cname{o_n}}^m \in \ibind{\subst}{\omega}$ if and only if $\tup{y_1,\ldots,y_n}^m \in \omega$, such that for every $i \in \set{1,\ldots,n}$, we have 
$\cname{o_i} = y_i$ if $y_i \in \dom_\types$, or $\cname{o_i} = \subst(y_i)$ if $y_i \in \vars$.
 For example, given $\omega=\set{\tup{x,y}^2,\tup{x,1}}$ and $\subst=\set{x\mapsto 1,y\mapsto 2}$, we have $\ibind{\subst}{\omega} = \set{\tup{1,2}^2,\tup{1,1}}$.

\begin{notion}[Transition enablement]
	Let $\dbn$ be a \dbnet with control layer $\tup{\places,\transitions,\inflow,\outflow,\rbflow,\coloring,\quass,\guass,\aass}$. A transition $t \in \transitions$ is \emph{enabled} in a $\dbn$-snapshot $\dbstate{\I}{\marking}$, written  $\enabled{\dbstate{\I}{\marking}}{t}{\mode}$, if:
	\begin{compactenum}
		\item for every place $p \in \places$, $m(p)$ provides enough  tokens matching those required by inscription $\omega = \inflow(\tup{p,t})$ once $\omega$ is bound by $\mode$,  i.e,  $\ibind{\mode}{\omega} \subseteq m(p)$;
		\item $\guass(t)\mode$ is true;
		\item $\mode$ is injective over $\freshvars{t}$, thus guaranteeing that fresh variables are assigned to pairwise distinct values by $\mode$, and for every fresh variable $\nu \in \freshvars{t}$, $\mode(\nu) \notin (\adom[\vartype(\nu)]{\I} \cup \adom[\vartype(\nu)]{m})$.
	\end{compactenum}
\end{notion}
\begin{notion}[Induced action instance]
Let $\cl = \langle\places,\transitions,\inflow,\outflow,\rbflow,$ $\coloring,\quass,\guass,\aass\rangle$ be a $\types$-typed control layer, and let $t \in \transitions$ be a transition of $\cl$ such that $\aass(t) = \tup{\alpha,\omega}$, with $\apar{\alpha} = \tup{x_1,\ldots,x_n}$ and $\omega = \tup{y_1,\ldots,y_n}$. The \emph{action instance induced} by transition $t \in \transitions$ under binding $\mode$, written $\indact{t}{\mode}$, is the action instance $\alpha\mode'$, where $\mode': \apar{\alpha} \rightarrow \dom_\types$ is a substitution for the formal parameters of $\alpha$, defined as: for every $i \in \set{1,\ldots,n}$, if $y_i \in \dom_\types$, then $\mode'(x_i) = y_i$; if instead $y_i \in \vars$, then $\mode'(x_i) = \subst(y_i)$.
\end{notion}
The firing of an enabled transition under some mode has then a threefold effect.
First, all tokens present in control places that are used to match the input inscriptions are consumed. Second, the action instance induced by the firing is applied on the current database instance. If such an action instance can be successfully applied, the database instance is updated accordingly; if not, the database instance is kept unaltered (thus realizing a rollback).
Third, tokens constructed using the inscriptions on output arcs are produced, and inserted into their target places, considering either normal output arcs or rollback arcs depending on whether the induced action instance is successfully applied or not.

\begin{notion}[Transition firing]
Let $ \dbn = \tup{\types,\pers,\dl,\cl}$ be a \dbnet with $\cl = \tup{\places,\transitions,\inflow,\outflow,\rbflow,\coloring,\quass,\guass,\aass}$, and $s_1=\dbstate{\I_1}{\marking_1}$, $s_2=\dbstate{\I_2}{\marking_2}$ be two $\dbn$-snapshots.
	Let $t \in \transitions$ be a transition of $\cl$, and $\mode$ be a binding for $t$, such that $\enabled{s_1}{t}{\mode}$. 	We say that $t$ \emph{fires} in $\state_1$ with binding $\sigma$ producing $\state_2$, written $\fire{\state_1}{t}{\mode}{\state_2}$, if the following conditions hold: given $\I_3 = \doact{\indact{t}{\mode}}{\I_1}$,
\begin{compactitem}[$\bullet$]
	\item if $\I_3$ is compliant with $\pers$, then $\I_2 = \I_3$, otherwise $\I_2 = \I_1$;
	\item For every control place $p \in \places$, given $\omega_{in} = \inflow(\tup{p,t})$, $\omega_{out} = \outflow(\tup{t,p})$, and $\omega_{rb} = \rbflow(\tup{t,p})$, we have 
	\[m_2(p) = (m_1(p) - \ibind{\mode}{\omega_{in}}) + k_{out} \cdot \ibind{\mode}{\omega_{out}} + (1-k_{out}) \cdot \ibind{\mode}{\omega_{rb}},\]
	where $k_{out} = 1$ if $\I_3$ is compliant with $\pers$, and $k_{out} = 0$ otherwise.
\end{compactitem}
\end{notion}

The execution semantics of a \dbnet is captured by a possibly \emph{infinite-state labeled transition system} (LTS) that accounts for all possible executions of the control layer starting from an initial snapshot. States of this transition systems are \dbnet snapshots, and transitions model the effect of firing \dbnet transitions under given bindings.
Formally, the execution semantics of a \dbnet $ \dbn = \tup{\types,\pers,\dl,\cl}$ with $\cl = \tup{\places,\transitions,\inflow,\outflow,\rbflow,\coloring,\quass,\guass,\aass}$ is given in terms of an LTS  $\tsys{\dbn} = \tup{\states,\istate,\trans}$, where:
\begin{compactitem}[$\bullet$]
\item $\states$ is a possibly infinite set of $\dbn$-snapshots;
\item $\istate$ is the \emph{initial $\dbn$-snapshot};
\item $\trans \subseteq \states \times \transitions \times \states$ is a \emph{transition relation} over states, labeled by transitions $\transitions$;
\item $\states$ and $\trans$ are defined by simultaneous induction as the smallest sets such that:
\begin{inparaenum}[\it (i)]
\item $\istate \in \states$;
\item given a $\dbn$-snapshot $\state \in \states$, for every transition $t \in \transitions$, binding $\mode$, and $\dbn$-snapshot $\state'$, if $\fire{\state}{t}{\mode}{\state'}$ then $\state' \in \states$ and $\state \overset{t}{\rightarrow}{\state'}$.
\end{inparaenum}
\end{compactitem}

\section{Discussion and Conclusion}
We believe that \dbnets have the potential of stimulating discussion, and possibly spawning new lines of investigation, for researchers interested in the foundations and applications of data-aware processes. We consider in particular the impact on modeling, verification, and simulation.

\smallskip\noindent{\bf Modeling.}
From the modeling point of view, \dbnets incorporate all typical abstractions needed in data-aware business processes.
In this light, our model covers all the distinctive features of various Petri net classes enriched with data, as well as those of data-centric processes. More formally, in terms of expressiveness, we observe/conjecture the following correspondences.
First of all, \dbnets subsume $\nu$-PNs \cite{RVFE11}, and become expressively equivalent to $\nu$-PNs when: 
\begin{inparaenum}[\it (i)]
\item there is only one unary color assigning places to the only one unordered countably infinite data type, 
\item the data logic layer is empty. 	
 \end{inparaenum}
 With such a restrictions, the only modeling construct not natively provided by $\nu$-PNs is that of arbitrary external input, which can be however simulated using $\nu$-PNs by following the strategy defined in \cite{AAA16}.
Second, \dbnets are expressively equivalent to recently introduced formal models for data-centric business processes, like DCDSs \cite{BCDD13} and DMSs \cite{AAA16}. To transform those models into a \dbnet, it is sufficient to realize a control layer that simulates the application of condition-action rules. The translation of a \dbnet into those models, instead, is more convoluted, but can be attacked by leveraging the technique introduced in \cite{MonR16} to encode $\nu$-PNs into DCDSs. Since DCDSs and DMSs are expressively equivalent to the richest models for business artifacts, such a correspondence paves the way towards the study of CPN-based business artifacts, making approaches like that of \cite{Loh13} truly data-aware.

However, we also stress that \dbnets go beyond the aforementioned approaches, since they conceptually componentize the different aspects of a dynamic system, giving first-class citizenships to relations, constraints, queries, database access points in the process, database updates triggered by the process, external inputs, and so on. This opens another interesting line of research, focused on understanding how to exploit \dbnets from the conceptual and methodological point of view, as well as on their exploitation to formalize concrete BP management systems like Bizagi, Bonita, and Camunda.

\smallskip\noindent
\textbf{Formal Verification.}
It is straightforward to see that \dbnets, in their full generality, are Turing-complete, and consequently that all standard verification tasks such as reachability, coverability, and model checking are undecidable.  However, the fact that the different aspects of a dynamic system are conceptually separated in \dbnets make them an ideal model to study in a fine-grained way how such aspects impact on undecidability and complexity of verification tasks, and how should they be controlled to guarantee decidability/tractability. For example, it is known from the literature that the presence of ordered vs. unordered data types, and of (globally) fresh inputs, is intimately connected to the boundaries of decidability for reachability \cite{Las16}. A similar observation holds for the presence of negation in the queries used to inspect the persistence layer, as well as for the arity of relation schemas contained therein \cite{AAA16}. Interestingly, \dbnets do not only provide a comprehensive model to fine-tune all such parameters, but also allow to study how they interact with each other. 

Within this space, we consider of particular importance the case where the \dbnet obeys to the so-called \emph{state-boundedness} property \cite{BCDD13,BCDM14,MonR16}. Intuitively, in the context of \dbnets, this means that the control layer is depth- and width-bounded \cite{RVFE11,MonR16}, and that the underlying database instance does not simultaneously employ more than a pre-defined number of elements (which however may arbitrarily change over time). Such an assumption still allows for the \dbnet to visit infinitely many different snapshots along its runs, as no restriction is imposed on the size of the type domains, from which external inputs are borrowed. It has been shown that model checking data-aware dynamic systems against first-order variants of $\mu$-calculus is indeed decidable, giving a constructive technique to carry out the verification task \cite{BCDD13,MonR16}. This opens up another interesting line of investigation on how to check, or guarantee using modeling strategies, that a \dbnet state-bounded, leveraging recent results \cite{RVFE11,BCDM14,MonC16,MonR16}.

Finally, \dbnets paves the way towards the formal analysis of additional properties, which only become relevant when CPNs are combined with relational databases. We mention in particular two families of properties. The first is related to \emph{rollbacks}, so as to check whether it is always (or never) the case that a transition induces a failing action. The second is related to the \emph{true concurrency} present in a \dbnet, which may contain transitions that appear to be concurrent by considering the control layer in isolation, but have instead to be sequenced due to the interplay with the persistence layer (and its constraints).

\smallskip
\noindent
\textbf{Simulation and Benchmarking.}
Since the control layer of \dbnets is grounded on CPNs, all simulation techniques developed for CPNs can be seamlessly lifted to our setting. The result of a \dbnet simulation produces, as a by-product, a final, database instance, populated through the execution of the control layer. On the one hand, this database instance may be scaled up at one's pleasure, by just changing the simulation parameters. On the other hand, the obtained database instance implicitly reflects the footprint of the control layer, which, e.g., inserts data in a certain order. This makes the obtained database instance much more intriguing than one synthetically generated without considering how data are generated over time. In this light, simulation of \dbnets has the potential of providing novel insights into the problem of data benchmarking \cite{LRXC15}, especially in the context of data preparation for process mining \cite{Aals13,CMSA15}. 

\bibliographystyle{splncs03}
\bibliography{mybib}

\end{document}